\documentclass[
  reprint,
  superscriptaddress,
  amsmath,amssymb,
  aps,pra,
  showkeys
]{revtex4-2}

\usepackage{graphicx}
\usepackage[dvipsnames]{xcolor}
\usepackage[textwidth=1.5cm]{todonotes}
\usepackage{dcolumn}
\usepackage{bm}
\usepackage{braket}
\usepackage[markup=underlined]{changes}
\usepackage[normalem]{ulem}
\usepackage{url}
\usepackage[nopatch=footnote]{microtype}

\newif\iftrackchanges

\newcommand{\DEACTIVE}{\trackchangesfalse}
\DEACTIVE

\DeclareRobustCommand{\macroremoved}[1]{%
  \iftrackchanges
    \textcolor{blue}{%
      \ifmmode
        \text{\sout{\ensuremath{#1}}}%
      \else
        \sout{#1}%
      \fi
    }%
  \else
  \fi
}

\newcommand{\FigRef}[1]{Fig.~\ref{#1}}

\makeatletter
\AtBeginDocument{\g@addto@macro\UrlBreaks{\do\-}}
\g@addto@macro\bibfont{\raggedright}
\def\@bibdataout@aps{%
 \immediate\write\@bibdataout{%
  @CONTROL{%
   apsrev42Control%
   \longbibliography@sw{%
    ,author="48",editor="1",pages="0",title="0",year="1"%
   }{%
    ,author="48",editor="1",pages="0",title="",year="1"%
   }%
  }%
 }%
 \if@filesw
  \immediate\write\@auxout{\string\citation{apsrev42Control}}%
 \fi
}%
\makeatother

\begin{document}

\title{Persistent structural regimes in Bitcoin transactions relate to market cycles}

\author{Bruno Araujo dos Santos}
\affiliation{Exponential Science Foundation, Lugano, Switzerland}

\author{Gabriele Di Antonio}
\thanks{These authors contributed equally to this work.\\ \url{gabriele.diantonio@upf.edu} (G.~Di Antonio),\\ \url{marcojavarone@gmail.com} (M.~A.~Javarone).}
\affiliation{Pompeu Fabra University, Barcelona, Spain}

\author{Marco Alberto Javarone}
\thanks{These authors contributed equally to this work.\\ \url{gabriele.diantonio@upf.edu} (G.~Di Antonio),\\ \url{marcojavarone@gmail.com} (M.~A.~Javarone).}
\affiliation{Exponential Science Foundation, Lugano, Switzerland}

\begin{abstract}
Bitcoin's blockchain is a public record of every on-chain transaction, while the price of bitcoin is set by trading on external markets. We ask whether on-chain activity alone reveals regimes associated with market trends, and whether this association remains stable over time. We summarise each week of blockchain activity from 2012 to 2026 using four measures of transaction structure and fit a two-state hidden semi-Markov model without price data. The sole market-informed constraint is an asymmetry in regime persistence, motivated by the observation that Bitcoin's price increases tend to outlast its declines. During earlier market cycles, the inferred regimes exhibit distinct transaction structures and align with periods of rising and falling prices more closely than random sequences with the same regime-duration distributions. The ongoing downturn, however, breaks this correspondence: the blockchain remains in the longer-lived regime, while its transaction structure resembles neither previous downturns nor previous upturns. Our findings support a relationship between on-chain activity and market dynamics, while highlighting its potentially time-dependent nature.
\end{abstract}

\maketitle

\section*{Introduction}

Bitcoin offers an unusual setting for studying the relation between economic
activity and market dynamics. Its blockchain provides a public historical
record of transactions, while BTC is simultaneously traded in external
markets where its price is formed. The transaction record makes it possible
to study how activity on the network changes over time rather than relying
only on market variables such as price and volume. Previous work has shown
that the Bitcoin transaction network has a non-trivial and evolving structure
\cite{Kondor2014Rich,Vallarano}, and that changes in this structure can
accompany substantial movements in the Bitcoin price
\cite{Kondor2014Interplay,liasons,Alberto_Javarone_2023,DiAntonio2023Rationality}. This raises the question of whether on-chain
activity itself can be organised into persistent regimes, and how such
regimes relate to the broader evolution of the Bitcoin market.

Regime-based descriptions have been used extensively to capture changes in
cryptocurrency market behaviour. Markov-switching models reveal distinct
volatility regimes in Bitcoin returns \cite{ARDIA2019266}, while hidden Markov
models have been used to identify unobserved bullish, stable, and bearish
price regimes \cite{Giudici2020} and broader positive and negative phases
across cryptocurrency markets \cite{Pennoni2022}. In these approaches,
however, the latent states are inferred primarily from prices, returns, or
volatility. A different question is whether persistent regimes can instead be
identified in the transaction activity recorded on the blockchain,
and compared with the market without including price in the observation
vector. This distinction makes it possible to ask not
only whether blockchain and market dynamics are related, but also whether
that relation remains stable over successive market cycles.

Here we study whether Bitcoin on-chain activity can be described by a small
number of persistent latent regimes and whether their relation to the market
remains stable over time. The observed asymmetry between longer periods of
overall price increase and shorter periods of overall decline (\FigRef{fig:temporal}) motivates a
two-timescale persistence hypothesis, which we represent with a two-state
hidden semi-Markov model (HSMM) fitted to multivariate transaction-structural
observables. Market price is not included in the model but it motivates the duration constraints and the choice between two solutions. Our question is therefore whether the transaction record admits a persistent decomposition related to the market, not whether a uniquely preferred segmentation emerges without market-informed choices. We assess the selected chronology against a duration-matched random benchmark and compare its transaction structure with price-defined phases.
Over the earlier market cycles, the selected regimes correspond substantially
to periods of price increase and decline (\FigRef{fig:states}) and exhibit distinct transaction profiles (\FigRef{fig:nonstationarity}). This correspondence is absent in the most recent downturn, where the blockchain remains in the more persistent regime and its configuration differs from previous periods of decline. Whether this departure from the historical correspondence is temporary or reflects a more persistent change in the relation between transaction structure and market dynamics remains open.

\section*{Materials and methods}

\subsection*{Market benchmark and temporal hypothesis}

Bitcoin's market history has commonly been described in terms of
multi-year boom--bust cycles occurring on a time scale comparable to the
approximately four-year reward halving schedule
\cite{math11030698,MBAKOB2024100178}.
Although the correspondence between halvings and market turning points is
neither exact nor necessarily causal
\cite{math11030698,LASHKARIPOUR2024106198}, the repeated cycle-scale structure
motivates a systematic description of Bitcoin's market phases.

For the market-side analysis, we use BTC/USD price data \cite{zielak_kaggle_bitcoin,ff137_bitstamp_repo,binance_public_data}. The daily price is taken as the final
available close of each UTC day, over the January 2012--August 2026
observation period.

\begin{figure}[ht!]
\centering
\includegraphics[width=\columnwidth]{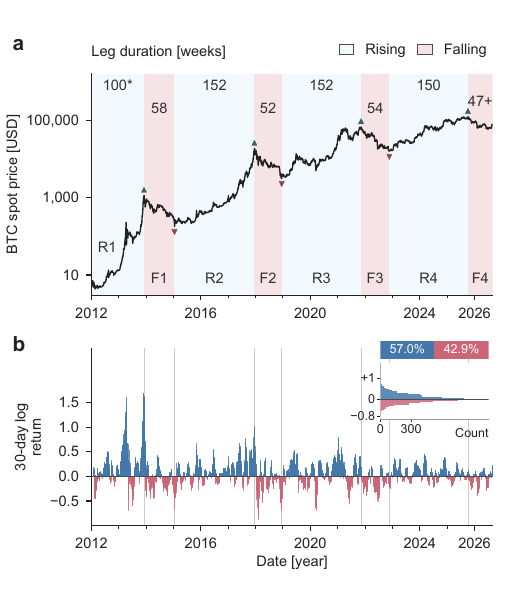}
\caption{\textbf{Temporal asymmetry of Bitcoin price dynamics.}
\textbf{a,} Bitcoin spot price [USD] on a logarithmic scale, with
alternating Rising and Falling legs defined by the major price peaks
and troughs. Triangles mark the extrema and labels give the corresponding
leg durations in weeks. The first Rising leg (*) is truncated by the sample
boundary and the final Falling leg (+) is ongoing.
\textbf{b,} Local price trend \(r_n\), the 30-day log return of the price.
Positive and negative values are shown separately, with vertical lines
marking the extrema from panel \textbf{a}. The inset gives the distribution
of \(r_n\); \(57.0\%\) of
valid days have positive trend and \(42.9\%\) negative trend.}
\label{fig:temporal}
\end{figure}

We define a retrospective market benchmark by dividing the price history
into alternating Rising and Falling phases.
The phases are separated by the major peaks and troughs of the price, which
we locate using the halving dates (28 November 2012, 9 July 2016, 11 May 2020
and 20 April 2024). After each halving, the peak is the highest daily close
in the following two years, and the trough is the lowest daily close between
that peak and the next halving. Each halving cycle therefore contains one
peak and one trough. A Rising phase runs from a trough to the next peak, and
a Falling phase from a peak to the next trough. This gives four Rising
phases, \(R1\)--\(R4\), and four Falling phases, \(F1\)--\(F4\)
(Fig.~\ref{fig:temporal}a). \(R1\) begins at the start of the data, so its
true start is unknown. \(F4\) begins at the last peak, in October 2025, and
is still ongoing because the next halving has not yet occurred. We use these
phases in all later comparisons. They describe past market history and are
not real-time signals.

The resulting benchmark exhibits a clear temporal asymmetry between Rising
and Falling market phases. Excluding the sample-truncated \(R1\) and the
ongoing \(F4\), the completed Rising phases \(R2\)--\(R4\) persist for
approximately \(150\)--\(152\) weeks, whereas the completed Falling phases
\(F1\)--\(F3\) last only \(52\)--\(58\) weeks
(Fig.~\ref{fig:temporal}a). Thus, at the cycle scale, Rising phases last much
longer than Falling phases. 

A complementary local measure shows a similar imbalance at shorter
time scales. With \(P_n\) the closing price on day \(n\), we use the 30-day
log return \(r_n=\ln(P_n/P_{n-30})\), which is positive when the price is
higher than 30 days earlier. Across the full sample, \(r_n\) is
positive on \(57.0\%\) of days and negative on \(42.9\%\)
(Fig.~\ref{fig:temporal}b), so upward local trends occupy more of the
observation period than downward ones. Together with the asymmetry observed in the
market-phase durations, this motivates a latent-state description with one
relatively long and one relatively short persistence scale.

\subsection*{On-chain representation and feature preprocessing}

To explore persistent regimes in Bitcoin transaction activity, we use blockchain data from January 2012 to August 2026, obtained from Google BigQuery
\cite{google_bigquery_bitcoin}.
Following previous works that describe
each block by its numbers of transactions, inputs and outputs
\cite{Alberto_Javarone_2023,DiAntonio2023Rationality}, we compute four
quantities for every block.

For block \(b\), let \(I_b\) and \(O_b\) be the total numbers of inputs and
outputs of its transactions, and \(N_b\) its number of transactions. We define
\begin{equation}
\begin{aligned}
B_b &=
\frac{I_b-O_b}{I_b+O_b+\epsilon},
\\
C_b &=
\ln\!\left(
1+\frac{I_b}{N_b+\epsilon}
\right),
\\
F_b &=
\ln\!\left(
1+\frac{O_b}{N_b+\epsilon}
\right),
\end{aligned}
\label{eq:block_features}
\end{equation}
where \(\epsilon=10^{-12}\) only prevents division by zero. The
\emph{input--output imbalance} \(B_b\) lies between \(-1\) and \(1\): it is
positive when the block has more inputs than outputs and negative when it has
more outputs than inputs. The \emph{consolidation} \(C_b\) and the
\emph{fan-out} \(F_b\) measure the average numbers of inputs and of outputs per
transaction, on a logarithmic scale. The fourth quantity is the
\emph{transaction entropy}
\begin{equation}
S_b=-\sum_{i:v_{b,i}>0}p_{b,i}\ln p_{b,i},
\qquad
p_{b,i}=\frac{v_{b,i}}{\sum_j v_{b,j}},
\label{eq:entropy}
\end{equation}
where \(v_{b,i}\) is the total amount of bitcoin in the outputs of transaction \(i\) in
block \(b\). The entropy is high when value is spread evenly over many
transactions and low when a few transactions carry most of it.

These quantities record only how many inputs and outputs transactions have and how much bitcoin they move. They do not tell why a transaction was made, for example whether it is a trade, a payment, or a transfer between a user's own addresses.

Values change strongly from block to block, so we group the blocks into
consecutive sets of 1008 blocks and take the median of each quantity within
each set. A block is added about every ten minutes, so each set covers about
one week (6.7 days on average). From now on we will refer to these sets simply as \emph{weeks}; they are the
time step of the whole analysis. We write \(y_{t,f}\) for the median of
quantity \(f\) in week \(t\).

Between 2012 and 2026, the typical values of imbalance, fan-out and entropy
change considerably. A model fitted directly to these values could simply
separate early years from later years, instead of finding states that recur.
We therefore replace each of these three values by its percentile rank in the
recent past: \(u_{t,f}\) is the fraction of the
previous 156 weeks (about three years) in which quantity \(f\) was lower
than in week \(t\). For example, \(u_{t,f}=0.9\) means that the value in
week \(t\) is higher than in 90\% of the previous three years. Formally,
\begin{equation}
u_{t,f}
=
\frac{1}{|\mathcal{W}_{t}^{-}|}
\sum_{t'\in\mathcal{W}_{t}^{-}}
\left[
\mathbb{I}(y_{t',f}<y_{t,f})
+\frac{1}{2}\mathbb{I}(y_{t',f}=y_{t,f})
\right],
\label{eq:percentile}
\end{equation}
where \(\mathcal{W}_{t}^{-}\) is the set of these previous weeks,
\(\mathbb{I}\) equals 1 when its condition holds and 0 otherwise. Near the
start of the data, fewer than 156 previous weeks exist and we use all of
them. We require at least 39 previous weeks (about nine months), so the first
39 weeks serve only as history and 760 weeks remain for the model.
Consolidation changes much less over the long run, so we use its weekly
median \(y_{t,f}\) directly.

Finally, we put the four series on a common scale. For each of them we
subtract its median over the 760 weeks and divide by its interquartile
range, which gives the model inputs \(x_{t,f}\). A value of zero therefore
means a typical week: for consolidation, a typical level; for imbalance,
fan-out and entropy, a typical percentile rank, not a typical raw value. For
each week \(t\), the model observes the four inputs
\begin{equation}
\mathbf{x}_t=
\left(
x_{t,\mathrm{con}},
x_{t,\mathrm{IO}},
x_{t,\mathrm{fan}},
x_{t,\mathrm{ent}}
\right)^{\!\top},
\label{eq:observation_vector}
\end{equation}
shown in Fig.~\ref{fig:features}.

\subsection*{Hidden semi-Markov model}
The market analysis (\FigRef{fig:temporal}) shows that the characteristic
duration of Rising phases is substantially longer than that of Falling
phases, motivating the consideration of two distinct persistence scales
in the latent blockchain dynamics. Consequently, we model the temporal
evolution of the on-chain observation sequence using a two-state hidden
semi-Markov model (HSMM).
A standard hidden Markov model (HMM) represents state
persistence through repeated self-transitions, implying a geometric
distribution of state dwell times \cite{Rabiner1989ATO,YU2010215}.
This is restrictive for the present problem because regime duration is
itself part of the hypothesis. An HSMM instead assigns an explicit
duration distribution to each latent state, allowing persistence to be
modelled separately from the transition process
\cite{YU2010215}.
We therefore consider two states, denoted Long (\(L\)) and Short (\(S\)),
which represent blockchain regimes with relatively long and short
persistence, respectively. These labels refer to temporal persistence,
not to Rising and Falling market phases or to trading positions.
Market price is not included in the HSMM observation vector; within the
model, it serves only to motivate the asymmetric duration structure. The state-dependent
feature distributions and decoded transition dates are inferred from
the on-chain observations for each fit. The model structure is
summarised in Fig.~\ref{fig:states}a.

Let \(z_t\in\{L,S\}\) denote the latent state in week \(t\). Conditional on
\(z_t\), the four model inputs are modelled by a diagonal Gaussian emission
distribution,
\begin{equation}
p(\mathbf{x}_t\mid z_t=s)
=
\prod_{f=1}^{4}
\mathcal{N}\!\left(
x_{t,f};\mu_{s,f},\sigma_{s,f}^{2}
\right),
\qquad s\in\{L,S\},
\label{eq:emissions}
\end{equation}
so that each state has its own mean and variance for each input. The
diagonal covariance is a simplifying assumption: given the state, the inputs
are treated as independent, although they are built from the same
transaction counts and can therefore be correlated.

Temporal persistence is represented separately through state-specific
duration distributions. An \emph{episode} is a period of consecutive weeks
in the same state, and its duration \(d\) is the number of weeks it
contains. For an episode in state \(s\), we take a normal random variable
\(X\) with mean \(\nu_s\) and standard deviation \(\tau_s\). Each duration
\(d\) in an allowed range \(\mathcal{D}_s\) then receives the probability
that \(X\) rounds to \(d\),
\begin{equation}
p_s(d)\;\propto\;\Pr\!\left(d-\tfrac12<X<d+\tfrac12\right),
\qquad d\in\mathcal{D}_s,
\label{eq:duration}
\end{equation}
rescaled so that these probabilities sum to one. Durations outside
\(\mathcal{D}_s\) have probability zero.

The Long/Short hypothesis enters through these duration distributions. The
standard deviation \(\tau_s\) and the allowed range \(\mathcal{D}_s\) are set
in advance. For the Long state, \(\tau_L=35\) weeks and
\(\mathcal{D}_L=\{75,\ldots,240\}\); for the Short
state, \(\tau_S=12\) weeks and \(\mathcal{D}_S=\{24,\ldots,85\}\). The mean \(\nu_s\) is fitted to the data, which gives
\(\nu_L=138.2\) weeks and \(\nu_S=66.8\) weeks. Because
durations outside \(\mathcal{D}_s\) are excluded, the mean and standard
deviation of the resulting durations can differ from \(\nu_s\) and
\(\tau_s\).

Self-transitions between consecutive episodes are excluded, following strict explicit-duration formulations
\cite{JohnsonWillsky2013}. Persistence is therefore represented by
Eq.~\eqref{eq:duration} rather than by repeated state self-transitions; with
two states, successive episodes alternate between \(L\) and \(S\).
The model is initialised by \(K\)-means clustering \cite{Lloyd1982} and
fitted using an EM-style forward--backward procedure, with posterior state
occupancies used to update the Gaussian parameters and posterior
episode-duration counts used to update \(\nu_L\) and \(\nu_S\). The decoded
chronology is obtained by explicit-duration Viterbi inference
\cite{YU2010215}.

We repeat the fit from twenty initialisations with different random seeds.
The runs converge to two distinct decoded segmentations: B1 is reached in
11 runs and contains eight episodes, whereas B2 is reached in nine runs
and contains seven (Fig.~\ref{fig:solutions}). B1 has the higher marginal
likelihood. We retain B2 for the market comparison because its states show a clearer correspondence with the market phases defined above. The null test described below (Chance agreement) accounts for this selection.

\subsection*{Comparison with the market phases}
We compare the decoded states with the market phases in four ways. First,
we measure the agreement between the state (Short or Long;
Fig.~\ref{fig:states}b) and the market direction (Falling or Rising;
Fig.~\ref{fig:temporal}a) over the 713 weeks before \(F4\), as the fraction
of weeks in which they match and as Cohen's \(\kappa\) \cite{Cohen1960}.
Cohen's \(\kappa\) corrects this fraction for the agreement expected by
chance, given how often each label occurs: \(\kappa=1\) means perfect
agreement, \(\kappa=0\) agreement at chance level, and \(\kappa<0\) less
agreement than chance. We compare \(\kappa\) with the null model described
in the next subsection. Second, for each episode we compute the annualised endpoint log-price slope,
\([\ln P_{\mathrm{end}}-\ln P_{\mathrm{start}}]/\Delta t\), where
\(P_{\mathrm{start}}\) and \(P_{\mathrm{end}}\) are the prices at the start
and end of the episode and \(\Delta t\) is the elapsed time in years;
this is the mean log return per year over the episode
(Fig.~\ref{fig:states}c; both fitted solutions in Fig.~\ref{fig:solutions}b).
Third, we take the median of each feature within each market phase, which
compares the phases with the transaction structure without the HSMM
(Fig.~\ref{fig:nonstationarity}b). Fourth, we summarise the
four-dimensional configuration with two orthogonal axes fitted to the
historical weeks only (\(R1\)--\(R4\) and \(F1\)--\(F3\)), with \(F4\) excluded from the axis
fitting. This exclusion does not extend to the full-sample preprocessing or
HSMM fitting. The first axis, LD1, is obtained by linear discriminant analysis (LDA) of Rising versus Falling weeks \cite{Fisher1936}: it is the weighted sum of the four inputs that best separates the two groups. Only the ratios between the weights matter, so we scale them so that their squares sum to one and choose their sign so that Falling weeks have higher LD1 values; the weights are shown in Fig.~\ref{fig:nonstationarity}c. The second axis, residual PC1, is the
leading principal component \cite{Jolliffe} of the historical weeks after
their LD1 component has been removed. All weeks, including those of \(F4\),
are then projected on these fixed axes (Fig.~\ref{fig:nonstationarity}d;
the \(R3\) phase in detail in Fig.~\ref{fig:R3}). LD1 is supervised by the historical
market labels and is used here to describe their contrast, not to establish
out-of-sample classification performance.

\subsection*{Chance agreement}

\begin{figure*}[ht!]
\centering
\includegraphics[width=\textwidth]{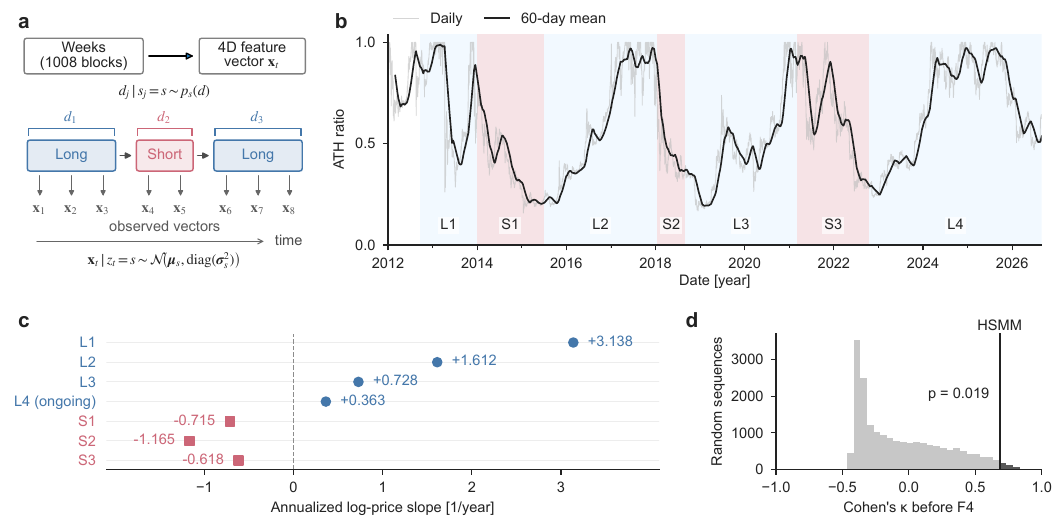}
\caption{\textbf{Latent blockchain regimes and Bitcoin market dynamics.}
\textbf{a,} Schematic of the two-state hidden semi-Markov model. Each
week (1008 blocks) is represented by a four-dimensional observed feature
vector \(\mathbf{x}_t\). Each episode \(j\) has a latent state \(s\in\{L,S\}\) and a
duration \(d_j\) drawn from the state-dependent distribution \(p_s(d)\).
During each state episode, successive observations \(\mathbf{x}_t\) are
generated according to
\(\mathbf{x}_t\mid z_t=s\sim
\mathcal{N}(\boldsymbol{\mu}_s,
\operatorname{diag}(\boldsymbol{\sigma}_s^2))\).
Self-transitions are excluded, so persistence is represented by the
explicit duration distributions.
\textbf{b,} Daily Bitcoin all-time-high ratio
\(R_n^{\mathrm{ATH}}=P_n/\max_{m\leq n}P_m\) shown in grey, together with
its trailing 60-day mean in black and the decoded Long and Short episodes
\(L1\)--\(L4\) and \(S1\)--\(S3\). The initial unshaded interval precedes
the first week used by the model. The latest price decline
occurs while the blockchain remains in \(L4\).
\textbf{c,} Annualised endpoint log-price slope
\([\ln P_{\mathrm{end}}-\ln P_{\mathrm{start}}]/\Delta t\) for each decoded
episode. All Long episodes have positive slopes and all Short episodes have
negative slopes.
\textbf{d,} Chance agreement. Grey histogram: Cohen's \(\kappa\) between the
market phases before \(F4\) and each of 20000 random state sequences with
episode durations drawn from the model's duration distributions. Vertical line: \(\kappa\) of the HSMM
states. Dark bars: random sequences that agree at least as well as the HSMM.}
\label{fig:states}
\end{figure*}

The HSMM states and the market phases both last for many weeks. Two such
slowly changing sequences can overlap over long stretches purely by chance,
so comparing them with labels drawn at random week by week would make the
agreement look more significant than it is. Instead, we generate 20000
random sequences of Long and Short episodes whose durations are drawn from
the model's duration distributions. Each random sequence covers the same 760 weeks as the data. Long
and Short episodes alternate, and the length of each episode is drawn from
\(p_L\) or \(p_S\) [Eq.~\eqref{eq:duration}], with the fitted values
\(\nu_L=138.2\) and \(\nu_S=66.8\). Each sequence starts at a random point
inside its first episode, so its transitions fall on random dates. No
blockchain data are used. For each random sequence we compute \(\kappa\) with
the market phases over the 713 weeks before \(F4\)
(Fig.~\ref{fig:states}d). The \(p\)-value is the fraction of random sequences
with \(\kappa\) at least as large as that of the HSMM. Since B2 was chosen
between two fitted solutions, we also report the \(p\)-value corrected for
this choice, \(2p\) (Bonferroni correction \cite{Dunn1961}), which holds
whatever the relation between the two solutions. This test takes the
duration settings as given;
it does not repeat the model fit or the choice between solutions.

\section*{Results}

\subsection*{Decoded chronology and market correspondence}

The selected segmentation resolves the on-chain sequence into seven
persistent episodes, \(L1\), \(S1\), \(L2\), \(S2\), \(L3\), \(S3\), \(L4\), spanning September 2012
to August 2026. Over most of the sample, the decoded blockchain chronology
follows the main Bitcoin market phases (Fig.~\ref{fig:states}b):
Short episodes occur predominantly during Falling phases, whereas Long
episodes occur predominantly during Rising phases.
The same relation is visible at the episode scale.
The annualised endpoint log-price slope is positive for
all four Long episodes and negative for all three Short episodes
(Fig.~\ref{fig:states}c).
The transition dates do not coincide exactly,
showing that the blockchain and price segmentations are related but not
identical. The clearest exception occurs in the ongoing 2025--2026 Falling
phase, during which every week so far is assigned to \(L4\).

Before \(F4\), the decoded state matches the market direction in \(88\%\)
of the 713 weeks, with Cohen's \(\kappa=0.68\). Only \(1.9\%\) of the
20000 random sequences in the duration-matched benchmark agree with the
market at least as well (\(p=0.019\); Fig.~\ref{fig:states}d). Because B2
was chosen from two fitted solutions, we also correct for this choice, which
gives \(p=0.039\).

\subsection*{Transaction structure of the latent regimes}
The two states differ clearly in transaction structure
(Fig.~\ref{fig:nonstationarity}a). Compared with the Long state, the Short
state has higher consolidation, input--output imbalance and fan-out, and
lower transaction entropy. Consolidation is compared with the whole sample:
in Short weeks, transactions have more inputs than in a typical week. The
other three features are percentile ranks, which compare each week with the
preceding three years. A positive imbalance in the Short state therefore
means a higher imbalance than usual at that time, not more inputs than
outputs; the same applies to fan-out and entropy. The open circles in
Fig.~\ref{fig:nonstationarity}a, which mark the medians of individual
episodes, show that this difference is not due to a single period. It is
clearest for consolidation, whose median is below zero in all four Long
episodes and above zero in all three Short episodes; the other features vary
more between episodes. In Short episodes, transactions therefore merge more
previously received coins and create more outputs than usual, while value is
concentrated in fewer transactions. An earlier study linked similar changes in
the input--output structure of blocks to periods of market stress
\cite{Alberto_Javarone_2023}.

\subsection*{Price-defined phases and the ongoing downturn}
\begin{figure*}[ht!]
\centering
\includegraphics[width=\textwidth]{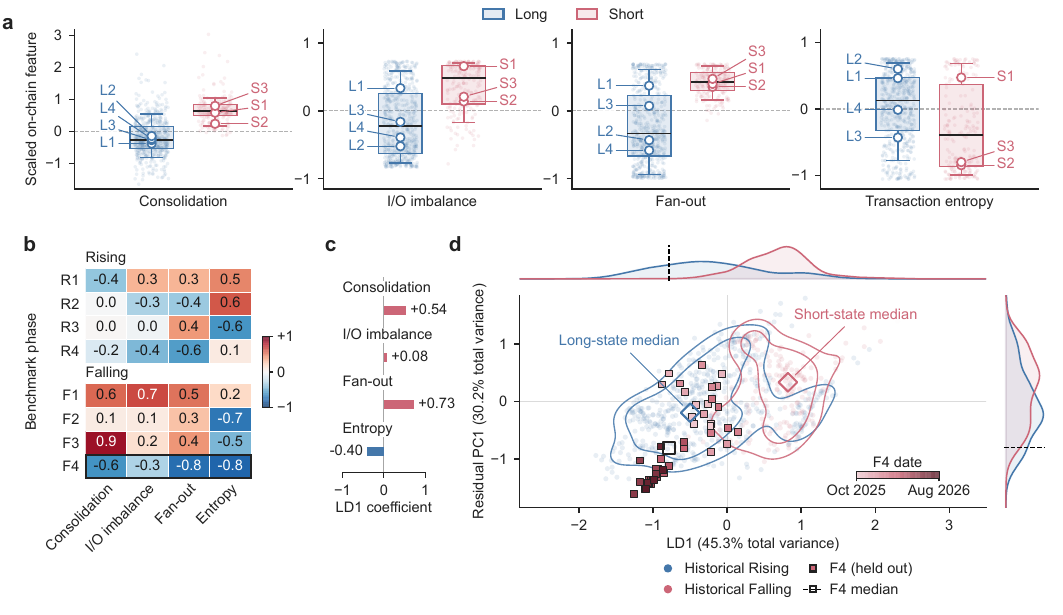}
\caption{\textbf{State anatomy and the changing on-chain signature of market phases.}
\textbf{a,} Distribution of the four transformed and scaled on-chain
features in the decoded Long and Short states. Points denote individual
weeks, boxes the interquartile range with median, whiskers the
10th--90th percentiles, and labelled open circles the medians of individual
episodes \(L1\)--\(L4\) and \(S1\)--\(S3\).
\textbf{b,} Median feature values in the externally defined Rising
(\(R1\)--\(R4\)) and Falling (\(F1\)--\(F4\)) market phases. The ongoing
\(F4\) phase is outlined.
\textbf{c,} Normalised coefficients of the linear-discriminant direction (LD1) fitted to historical Rising and Falling observations. Fan-out and consolidation
carry the largest positive weights, while transaction entropy enters with
negative weight.
\textbf{d,} Historical Rising and Falling observations projected onto
LD1 and the leading principal component
of the feature variation remaining after removal of LD1. The discriminant
and residual component are fitted using \(R1\)--\(R4\) and \(F1\)--\(F3\);
\(F4\) is excluded from the fit and projected afterwards. Marginal densities and contours show the historical Rising and Falling distributions. Small squares show the \(F4\) weeks, coloured by date from light (October 2025) to dark (August 2026). Diamonds mark the pooled Long- and Short-state medians, and the outlined square the median of \(F4\).
}
\label{fig:nonstationarity}
\end{figure*}

The market phases, defined from the price alone as in \FigRef{fig:temporal}a, allow the same comparison
without the HSMM (Fig.~\ref{fig:nonstationarity}b). In each of the three
completed Falling phases, consolidation, input--output imbalance and fan-out
are positive. The Rising phases are less uniform: consolidation remains close
to or below zero, whereas imbalance, fan-out and entropy change sign from one
phase to the next. This mirrors the HSMM result
(Fig.~\ref{fig:nonstationarity}a): the Falling phases share the signature of
the Short state, the Rising phases are as variable as the Long episodes, and
consolidation is again the most consistent feature.

Figure~\ref{fig:nonstationarity}c condenses this pattern into a single
direction. LDA, fitted to the weeks of the completed phases \(R1\)--\(R4\) and
\(F1\)--\(F3\) with \(F4\) left out, finds the weighted combination of the
four features, LD1, that best separates Rising from Falling weeks (see Materials and methods). Its
weights restate what Fig.~\ref{fig:nonstationarity}b shows: fan-out and
consolidation, positive in every Falling phase, carry the largest positive
weights, and entropy, low in two of the three Falling phases, enters with a
negative weight. Imbalance is also positive in every Falling phase but
receives little weight: it is built from the same input and output counts as
consolidation and fan-out, and adds little information beyond them.
Figure~\ref{fig:nonstationarity}d shows every week on LD1 and on residual
PC1, the direction of largest remaining variation, both fitted to the historical weeks alone. The historical Rising and Falling weeks form two overlapping but distinct distributions, and the Long- and Short-state medians of the HSMM (blue and red diamonds) lie inside them, respectively.

The ongoing Falling phase \(F4\) breaks the historical pattern. Its
consolidation, imbalance and fan-out are all negative, opposite in sign to
every earlier Falling phase, and its low entropy is shared with \(F2\) and
\(F3\) (Fig.~\ref{fig:nonstationarity}b). In
Fig.~\ref{fig:nonstationarity}d, the early \(F4\) weeks (light squares) fall within the historical Rising distribution, near the Long-state median,
consistent with the blockchain remaining in \(L4\). Later weeks (darker
squares) drift towards lower values on both axes, and many of the most recent ones
lie below every historical week along residual PC1. The current downturn
therefore lacks the transaction structure of past declines without matching
past Rising phases. Earlier mismatches between the price and blockchain
chronologies were offsets in timing, such as the gradual shift from \(L3\) to
\(S3\) within \(R3\) (Fig.~\ref{fig:R3}); the present one involves a feature
configuration unlike that of any earlier phase.

\section*{Discussion and conclusions}
This study asked whether Bitcoin transaction structure is organised into
persistent regimes and whether their relation to the market remains stable
over time. Over the historical market cycles, the transaction record admits
a persistent decomposition closely related to the market. Unlike regime
models fitted to prices, returns or volatility
\cite{ARDIA2019266,Giudici2020,Pennoni2022}, the HSMM observes transaction
structure alone, yet its Long and Short regimes follow the Rising and
Falling phases of the market. Three complementary observations support this
correspondence: the decoded chronology (Fig.~\ref{fig:states}b), the signs
of the episode price slopes (Fig.~\ref{fig:states}c), and a transaction
structure shared by the Short state and the price-defined Falling phases,
which emerges even without the model (Fig.~\ref{fig:nonstationarity}a,b).
Random sequences drawn from the same duration distributions rarely reach a
comparable agreement with the market (Fig.~\ref{fig:states}d).

The Short state admits a direct interpretation in terms of how bitcoin is
moved: transactions combine more previously received coins and create more
outputs than usual, while value is concentrated in fewer transactions. This
profile is consistent with earlier evidence that on-chain activity varies
with market conditions. Javarone et al.\ reported changes in input--output
structure around periods of market stress \cite{Alberto_Javarone_2023}, and
Di Antonio et al.\ found that an entropy measure of block structure dropped
sharply as Bitcoin first gained value \cite{DiAntonio2023Rationality}.
Bovet et al.\ found that blockchain activity becomes more
heterogeneous during price declines \cite{liasons}. Hoang and Baur
associated negative price shocks and elevated volatility with transfers from
private wallets to exchanges \cite{HOANG2022106622}, a mechanism that may
contribute to the Short-state profile. Our analysis does not identify the
actors or motives behind individual transactions; rather, it shows that a
small set of aggregate features combines into persistent configurations
that recur across market cycles.

The ongoing downturn \(F4\) departs from this pattern in a way that differs
from earlier discrepancies. Before \(F4\), the two chronologies disagreed
only on timing: during \(R3\), for instance, the weeks shifted gradually
towards the Short side of LD1, and the transition to \(S3\) preceded the end
of the Rising phase (Fig.~\ref{fig:R3}). In \(F4\), by contrast, the
blockchain remains in \(L4\) while the market declines; consolidation,
imbalance and fan-out are negative, unlike in every earlier Falling phase
(Fig.~\ref{fig:nonstationarity}b); and the later weeks move away from both
historical distributions (Fig.~\ref{fig:nonstationarity}d). The transaction
structure itself has changed, not merely the timing of the transitions. So
far, therefore, the relation between transaction structure and market
direction has not remained stable. As \(F4\) is still unfolding, whether
this departure is transient or lasting remains open.

Several developments in the Bitcoin ecosystem may have weakened the
historical link. On the blockchain side, Ordinals and inscriptions
introduced new uses of block space and have been shown to affect transaction
fees, block-space demand and overall network activity
\cite{BERTUCCI2024102338,wiedenmann2025ordinals}. Layer-2 systems such as
the Lightning Network allow payments to be made without each being recorded
on-chain, and their adoption has been associated with reduced congestion of
the base layer \cite{DIVAKARUNI2023103480}. Both developments alter which
activity reaches the blockchain, and hence what our features capture. The
market has evolved as well: derivatives contribute substantially to Bitcoin
price discovery \cite{ALEXANDER2020100776}, and recent work identifies an
important role for spot Bitcoin ETFs \cite{Mohamad2025}. As the ecosystem
matures, the blockchain and the market need not remain as closely coupled as
in earlier cycles. These mechanisms are plausible, but the cited studies do
not examine the features used here, and our analysis cannot attribute the
behaviour of \(F4\) to any of them.

Some limitations should be noted. The historical record contains only three
completed Falling phases, and consecutive weeks are serially dependent, so
the correspondence rests on a small number of independent repetitions.
Market history informs two modelling choices, the allowed durations of the
two states and the selection of B2 over B1, which attains the higher
likelihood (Fig.~\ref{fig:solutions}); our result is therefore the
existence of a market-related decomposition of the transaction record, not
a unique segmentation. The model is also deliberately simple, with two
states and Gaussian emissions that only approximate the bounded percentile
ranks. Finally, the analysis is descriptive: it neither tests causal
relations between blockchain activity and price, for which dedicated
methods exist \cite{Granger1969,Pearl2009,Sugihara2012,DiAntonio2025Causality}
and have been applied to Bitcoin \cite{liasons,Alberto_Javarone_2023}, nor
provides a trading signal.

Within these limits, our results show that Bitcoin's public transaction
record carries a persistent, market-related structure that can be
identified without including prices in the model. On-chain activity
therefore offers a view of market regimes that complements those based on
prices, returns and volatility. The correspondence, however, has not
carried over to the current downturn, which suggests that the relation
between Bitcoin's blockchain and its market evolves with the ecosystem
rather than remaining fixed.

\section*{Data availability}
Block and transaction data are available from the Google BigQuery public
dataset \url{bigquery-public-data.crypto_bitcoin} \cite{google_bigquery_bitcoin}.
Daily prices are Bitstamp BTC/USD closes for 2012--2024
\cite{zielak_kaggle_bitcoin,ff137_bitstamp_repo} and Binance BTC/USDT closes
for 2025--2026 \cite{binance_public_data}.

\section*{CRediT authorship contribution statement}
BADS: Conceptualization, Formal analysis,
Software, Writing -- original draft. GDA: Conceptualization, Formal analysis,
Supervision, Writing -- review \& editing. MAJ: Conceptualization, Supervision,
Writing -- review \& editing.

\section*{Declaration of competing interest}
The authors declare no competing interests.

\section*{Declaration of generative AI and AI-assisted technologies in the
manuscript preparation process}
During the preparation of this work the authors used large language models,
including Claude (Anthropic) and ChatGPT (OpenAI), to revise the manuscript text and to check and
extend the analysis code. After using these tools, the authors reviewed and
edited the content as needed and take full responsibility for the content of
the published article.

\begin{acknowledgments}
BADS gratefully acknowledges financial support from the Exponential Science Foundation.
\end{acknowledgments}

\makeatletter
\def\bibsection{\par\begingroup\baselineskip26\p@\bib@device{\hsize}{72\p@}%
\endgroup\nobreak\@nobreaktrue\addvspace{19\p@}}
\makeatother
\bibliography{RefLibrary}

\begin{thebibliography}{34}%
\makeatletter
\providecommand \@ifxundefined [1]{%
 \@ifx{#1\undefined}
}%
\providecommand \@ifnum [1]{%
 \ifnum #1\expandafter \@firstoftwo
 \else \expandafter \@secondoftwo
 \fi
}%
\providecommand \@ifx [1]{%
 \ifx #1\expandafter \@firstoftwo
 \else \expandafter \@secondoftwo
 \fi
}%
\providecommand \natexlab [1]{#1}%
\providecommand \enquote  [1]{``#1''}%
\providecommand \bibnamefont  [1]{#1}%
\providecommand \bibfnamefont [1]{#1}%
\providecommand \citenamefont [1]{#1}%
\providecommand \href@noop [0]{\@secondoftwo}%
\providecommand \href [0]{\begingroup \@sanitize@url \@href}%
\providecommand \@href[1]{\@@startlink{#1}\@@href}%
\providecommand \@@href[1]{\endgroup#1\@@endlink}%
\providecommand \@sanitize@url [0]{\catcode `\\12\catcode `\$12\catcode
  `\&12\catcode `\#12\catcode `\^12\catcode `\_12\catcode `\%12\relax}%
\providecommand \@@startlink[1]{}%
\providecommand \@@endlink[0]{}%
\providecommand \url  [0]{\begingroup\@sanitize@url \@url }%
\providecommand \@url [1]{\endgroup\@href {#1}{\urlprefix }}%
\providecommand \urlprefix  [0]{URL }%
\providecommand \Eprint [0]{\href }%
\providecommand \doibase [0]{https://doi.org/}%
\providecommand \selectlanguage [0]{\@gobble}%
\providecommand \bibinfo  [0]{\@secondoftwo}%
\providecommand \bibfield  [0]{\@secondoftwo}%
\providecommand \translation [1]{[#1]}%
\providecommand \BibitemOpen [0]{}%
\providecommand \bibitemStop [0]{}%
\providecommand \bibitemNoStop [0]{.\EOS\space}%
\providecommand \EOS [0]{\spacefactor3000\relax}%
\providecommand \BibitemShut  [1]{\csname bibitem#1\endcsname}%
\let\auto@bib@innerbib\@empty
\bibitem [{\citenamefont {Kondor}\ \emph
  {et~al.}(2014{\natexlab{a}})\citenamefont {Kondor}, \citenamefont {Pósfai},
  \citenamefont {Csabai},\ and\ \citenamefont {Vattay}}]{Kondor2014Rich}%
  \BibitemOpen
  \bibfield  {author} {\bibinfo {author} {\bibfnamefont {D.}~\bibnamefont
  {Kondor}}, \bibinfo {author} {\bibfnamefont {M.}~\bibnamefont {Pósfai}},
  \bibinfo {author} {\bibfnamefont {I.}~\bibnamefont {Csabai}},\ and\ \bibinfo
  {author} {\bibfnamefont {G.}~\bibnamefont {Vattay}},\ }\bibfield  {title}
  {\bibinfo {title} {Do the rich get richer? {An} empirical analysis of the
  {Bitcoin} transaction network},\ }\href
  {https://doi.org/10.1371/journal.pone.0086197} {\bibfield  {journal}
  {\bibinfo  {journal} {PLOS ONE}\ }\textbf {\bibinfo {volume} {9}},\ \bibinfo
  {pages} {e86197} (\bibinfo {year} {2014}{\natexlab{a}})}\BibitemShut
  {NoStop}%
\bibitem [{\citenamefont {Vallarano}\ \emph {et~al.}(2020)\citenamefont
  {Vallarano}, \citenamefont {Tessone},\ and\ \citenamefont
  {Squartini}}]{Vallarano}%
  \BibitemOpen
  \bibfield  {author} {\bibinfo {author} {\bibfnamefont {N.}~\bibnamefont
  {Vallarano}}, \bibinfo {author} {\bibfnamefont {C.~J.}\ \bibnamefont
  {Tessone}},\ and\ \bibinfo {author} {\bibfnamefont {T.}~\bibnamefont
  {Squartini}},\ }\bibfield  {title} {\bibinfo {title} {{Bitcoin} transaction
  networks: An overview of recent results},\ }\href
  {https://doi.org/10.3389/fphy.2020.00286} {\bibfield  {journal} {\bibinfo
  {journal} {Frontiers in Physics}\ }\textbf {\bibinfo {volume} {8}},\ \bibinfo
  {pages} {286} (\bibinfo {year} {2020})}\BibitemShut {NoStop}%
\bibitem [{\citenamefont {Kondor}\ \emph
  {et~al.}(2014{\natexlab{b}})\citenamefont {Kondor}, \citenamefont {Csabai},
  \citenamefont {Sz{\"u}le}, \citenamefont {P{\'o}sfai},\ and\ \citenamefont
  {Vattay}}]{Kondor2014Interplay}%
  \BibitemOpen
  \bibfield  {author} {\bibinfo {author} {\bibfnamefont {D.}~\bibnamefont
  {Kondor}}, \bibinfo {author} {\bibfnamefont {I.}~\bibnamefont {Csabai}},
  \bibinfo {author} {\bibfnamefont {J.}~\bibnamefont {Sz{\"u}le}}, \bibinfo
  {author} {\bibfnamefont {M.}~\bibnamefont {P{\'o}sfai}},\ and\ \bibinfo
  {author} {\bibfnamefont {G.}~\bibnamefont {Vattay}},\ }\bibfield  {title}
  {\bibinfo {title} {Inferring the interplay between network structure and
  market effects in {Bitcoin}},\ }\href
  {https://doi.org/10.1088/1367-2630/16/12/125003} {\bibfield  {journal}
  {\bibinfo  {journal} {New Journal of Physics}\ }\textbf {\bibinfo {volume}
  {16}},\ \bibinfo {pages} {125003} (\bibinfo {year}
  {2014}{\natexlab{b}})}\BibitemShut {NoStop}%
\bibitem [{\citenamefont {Bovet}\ \emph {et~al.}(2023)\citenamefont {Bovet},
  \citenamefont {Campajola}, \citenamefont {Mottes}, \citenamefont {Restocchi},
  \citenamefont {Vallarano}, \citenamefont {Squartini},\ and\ \citenamefont
  {Tessone}}]{liasons}%
  \BibitemOpen
  \bibfield  {author} {\bibinfo {author} {\bibfnamefont {A.}~\bibnamefont
  {Bovet}}, \bibinfo {author} {\bibfnamefont {C.}~\bibnamefont {Campajola}},
  \bibinfo {author} {\bibfnamefont {F.}~\bibnamefont {Mottes}}, \bibinfo
  {author} {\bibfnamefont {V.}~\bibnamefont {Restocchi}}, \bibinfo {author}
  {\bibfnamefont {N.}~\bibnamefont {Vallarano}}, \bibinfo {author}
  {\bibfnamefont {T.}~\bibnamefont {Squartini}},\ and\ \bibinfo {author}
  {\bibfnamefont {C.~J.}\ \bibnamefont {Tessone}},\ }\bibfield  {title}
  {\bibinfo {title} {The evolving liaisons between the transaction networks of
  {Bitcoin} and its price dynamics},\ }in\ \href
  {https://doi.org/10.7566/JPSCP.40.011002} {\emph {\bibinfo {booktitle}
  {Proceedings of Blockchain Kaigi 2022 (BCK22)}}},\ \bibinfo {series} {JPS
  Conference Proceedings}, Vol.~\bibinfo {volume} {40}\ (\bibinfo  {publisher}
  {The Physical Society of Japan},\ \bibinfo {year} {2023})\ p.\ \bibinfo
  {pages} {011002}\BibitemShut {NoStop}%
\bibitem [{\citenamefont {Javarone}\ \emph {et~al.}(2023)\citenamefont
  {Javarone}, \citenamefont {Di~Antonio}, \citenamefont {Vinci}, \citenamefont
  {Cristodaro}, \citenamefont {Tessone},\ and\ \citenamefont
  {Pietronero}}]{Alberto_Javarone_2023}%
  \BibitemOpen
  \bibfield  {author} {\bibinfo {author} {\bibfnamefont {M.~A.}\ \bibnamefont
  {Javarone}}, \bibinfo {author} {\bibfnamefont {G.}~\bibnamefont
  {Di~Antonio}}, \bibinfo {author} {\bibfnamefont {G.~V.}\ \bibnamefont
  {Vinci}}, \bibinfo {author} {\bibfnamefont {R.}~\bibnamefont {Cristodaro}},
  \bibinfo {author} {\bibfnamefont {C.~J.}\ \bibnamefont {Tessone}},\ and\
  \bibinfo {author} {\bibfnamefont {L.}~\bibnamefont {Pietronero}},\ }\bibfield
   {title} {\bibinfo {title} {Disorder unleashes panic in {Bitcoin} dynamics},\
  }\href {https://doi.org/10.1088/2632-072X/ad00f7} {\bibfield  {journal}
  {\bibinfo  {journal} {Journal of Physics: Complexity}\ }\textbf {\bibinfo
  {volume} {4}},\ \bibinfo {pages} {045002} (\bibinfo {year}
  {2023})}\BibitemShut {NoStop}%
\bibitem [{\citenamefont {Di~Antonio}\ \emph {et~al.}(2023)\citenamefont
  {Di~Antonio}, \citenamefont {Vinci}, \citenamefont {Pietronero},\ and\
  \citenamefont {Javarone}}]{DiAntonio2023Rationality}%
  \BibitemOpen
  \bibfield  {author} {\bibinfo {author} {\bibfnamefont {G.}~\bibnamefont
  {Di~Antonio}}, \bibinfo {author} {\bibfnamefont {G.~V.}\ \bibnamefont
  {Vinci}}, \bibinfo {author} {\bibfnamefont {L.}~\bibnamefont {Pietronero}},\
  and\ \bibinfo {author} {\bibfnamefont {M.~A.}\ \bibnamefont {Javarone}},\
  }\bibfield  {title} {\bibinfo {title} {The rise of rationality in blockchain
  dynamics},\ }\href {https://doi.org/10.1088/1367-2630/ad149a} {\bibfield
  {journal} {\bibinfo  {journal} {New Journal of Physics}\ }\textbf {\bibinfo
  {volume} {25}},\ \bibinfo {pages} {123042} (\bibinfo {year}
  {2023})}\BibitemShut {NoStop}%
\bibitem [{\citenamefont {Ardia}\ \emph {et~al.}(2019)\citenamefont {Ardia},
  \citenamefont {Bluteau},\ and\ \citenamefont {Rüede}}]{ARDIA2019266}%
  \BibitemOpen
  \bibfield  {author} {\bibinfo {author} {\bibfnamefont {D.}~\bibnamefont
  {Ardia}}, \bibinfo {author} {\bibfnamefont {K.}~\bibnamefont {Bluteau}},\
  and\ \bibinfo {author} {\bibfnamefont {M.}~\bibnamefont {Rüede}},\
  }\bibfield  {title} {\bibinfo {title} {Regime changes in {Bitcoin} {GARCH}
  volatility dynamics},\ }\href {https://doi.org/10.1016/j.frl.2018.08.009}
  {\bibfield  {journal} {\bibinfo  {journal} {Finance Research Letters}\
  }\textbf {\bibinfo {volume} {29}},\ \bibinfo {pages} {266} (\bibinfo {year}
  {2019})}\BibitemShut {NoStop}%
\bibitem [{\citenamefont {Giudici}\ and\ \citenamefont
  {Abu~Hashish}(2020)}]{Giudici2020}%
  \BibitemOpen
  \bibfield  {author} {\bibinfo {author} {\bibfnamefont {P.}~\bibnamefont
  {Giudici}}\ and\ \bibinfo {author} {\bibfnamefont {I.}~\bibnamefont
  {Abu~Hashish}},\ }\bibfield  {title} {\bibinfo {title} {A hidden {Markov}
  model to detect regime changes in cryptoasset markets},\ }\href
  {https://doi.org/10.1002/qre.2673} {\bibfield  {journal} {\bibinfo  {journal}
  {Quality and Reliability Engineering International}\ }\textbf {\bibinfo
  {volume} {36}},\ \bibinfo {pages} {2057} (\bibinfo {year}
  {2020})}\BibitemShut {NoStop}%
\bibitem [{\citenamefont {Pennoni}\ \emph {et~al.}(2022)\citenamefont
  {Pennoni}, \citenamefont {Bartolucci}, \citenamefont {Forte},\ and\
  \citenamefont {Ametrano}}]{Pennoni2022}%
  \BibitemOpen
  \bibfield  {author} {\bibinfo {author} {\bibfnamefont {F.}~\bibnamefont
  {Pennoni}}, \bibinfo {author} {\bibfnamefont {F.}~\bibnamefont {Bartolucci}},
  \bibinfo {author} {\bibfnamefont {G.}~\bibnamefont {Forte}},\ and\ \bibinfo
  {author} {\bibfnamefont {F.}~\bibnamefont {Ametrano}},\ }\bibfield  {title}
  {\bibinfo {title} {Exploring the dependencies among main cryptocurrency
  log-returns: A hidden {Markov} model},\ }\href
  {https://doi.org/10.1111/ecno.12193} {\bibfield  {journal} {\bibinfo
  {journal} {Economic Notes}\ }\textbf {\bibinfo {volume} {51}},\ \bibinfo
  {pages} {e12193} (\bibinfo {year} {2022})}\BibitemShut {NoStop}%
\bibitem [{\citenamefont {Chan}\ \emph {et~al.}(2023)\citenamefont {Chan},
  \citenamefont {Phoong}, \citenamefont {Phoong}, \citenamefont {Cheng},\ and\
  \citenamefont {Chen}}]{math11030698}%
  \BibitemOpen
  \bibfield  {author} {\bibinfo {author} {\bibfnamefont {J.~Y.-L.}\
  \bibnamefont {Chan}}, \bibinfo {author} {\bibfnamefont {S.~W.}\ \bibnamefont
  {Phoong}}, \bibinfo {author} {\bibfnamefont {S.~Y.}\ \bibnamefont {Phoong}},
  \bibinfo {author} {\bibfnamefont {W.~K.}\ \bibnamefont {Cheng}},\ and\
  \bibinfo {author} {\bibfnamefont {Y.-L.}\ \bibnamefont {Chen}},\ }\bibfield
  {title} {\bibinfo {title} {The {Bitcoin} halving cycle volatility dynamics
  and safe haven-hedge properties: A {MSGARCH} approach},\ }\href
  {https://doi.org/10.3390/math11030698} {\bibfield  {journal} {\bibinfo
  {journal} {Mathematics}\ }\textbf {\bibinfo {volume} {11}},\ \bibinfo {pages}
  {698} (\bibinfo {year} {2023})}\BibitemShut {NoStop}%
\bibitem [{\citenamefont {M'bakob}(2024)}]{MBAKOB2024100178}%
  \BibitemOpen
  \bibfield  {author} {\bibinfo {author} {\bibfnamefont {G.~B.}\ \bibnamefont
  {M'bakob}},\ }\bibfield  {title} {\bibinfo {title} {Bubbles in {Bitcoin} and
  {Ethereum}: The role of halving in the formation of super cycles},\ }\href
  {https://doi.org/10.1016/j.sftr.2024.100178} {\bibfield  {journal} {\bibinfo
  {journal} {Sustainable Futures}\ }\textbf {\bibinfo {volume} {7}},\ \bibinfo
  {pages} {100178} (\bibinfo {year} {2024})}\BibitemShut {NoStop}%
\bibitem [{\citenamefont {Lashkaripour}(2024)}]{LASHKARIPOUR2024106198}%
  \BibitemOpen
  \bibfield  {author} {\bibinfo {author} {\bibfnamefont {M.}~\bibnamefont
  {Lashkaripour}},\ }\bibfield  {title} {\bibinfo {title} {Some stylized facts
  about {Bitcoin} halving},\ }\href {https://doi.org/10.1016/j.frl.2024.106198}
  {\bibfield  {journal} {\bibinfo  {journal} {Finance Research Letters}\
  }\textbf {\bibinfo {volume} {69}},\ \bibinfo {pages} {106198} (\bibinfo
  {year} {2024})}\BibitemShut {NoStop}%
\bibitem [{\citenamefont {{Zielak
  (mczielinski)}}(2025)}]{zielak_kaggle_bitcoin}%
  \BibitemOpen
  \bibfield  {author} {\bibinfo {author} {\bibnamefont {{Zielak
  (mczielinski)}}},\ }\href@noop {} {\bibinfo {title} {{Bitcoin Historical
  Data}}},\ \bibinfo {howpublished} {Kaggle dataset,
  \url{https://www.kaggle.com/datasets/mczielinski/bitcoin-historical-data}}
  (\bibinfo {year} {2025}),\ \bibinfo {note} {licensed under CC BY-SA
  4.0}\BibitemShut {NoStop}%
\bibitem [{\citenamefont {de~Beer}(2025)}]{ff137_bitstamp_repo}%
  \BibitemOpen
  \bibfield  {author} {\bibinfo {author} {\bibfnamefont {M.}~\bibnamefont
  {de~Beer}},\ }\href@noop {} {\bibinfo {title}
  {bitstamp-btcusd-minute-data}},\ \bibinfo {howpublished}
  {\url{https://github.com/ff137/bitstamp-btcusd-minute-data}} (\bibinfo {year}
  {2025}),\ \bibinfo {note} {{GitHub} repository, accessed:
  2026-03-15}\BibitemShut {NoStop}%
\bibitem [{\citenamefont {{Binance}}(2026)}]{binance_public_data}%
  \BibitemOpen
  \bibfield  {author} {\bibinfo {author} {\bibnamefont {{Binance}}},\
  }\href@noop {} {\bibinfo {title} {{Binance Public Data}: spot klines,
  {BTCUSDT}, 1 hour}},\ \bibinfo {howpublished}
  {\url{https://data.binance.vision}} (\bibinfo {year} {2026})\BibitemShut
  {NoStop}%
\bibitem [{\citenamefont {Day}\ and\ \citenamefont
  {Medvedev}(2019)}]{google_bigquery_bitcoin}%
  \BibitemOpen
  \bibfield  {author} {\bibinfo {author} {\bibfnamefont {A.}~\bibnamefont
  {Day}}\ and\ \bibinfo {author} {\bibfnamefont {E.}~\bibnamefont {Medvedev}},\
  }\href@noop {} {\bibinfo {title} {Introducing six new cryptocurrencies in
  {BigQuery} public datasets---and how to analyze them}},\ \bibinfo
  {howpublished} {Google Cloud Blog,
  \url{https://cloud.google.com/blog/products/data-analytics/introducing-six-new-cryptocurrencies-in-bigquery-public-datasets-and-how-to-analyze-them}}
  (\bibinfo {year} {2019}),\ \bibinfo {note} {{Bitcoin} dataset
  \url{bigquery-public-data.crypto_bitcoin}}\BibitemShut {NoStop}%
\bibitem [{\citenamefont {Rabiner}(1989)}]{Rabiner1989ATO}%
  \BibitemOpen
  \bibfield  {author} {\bibinfo {author} {\bibfnamefont {L.~R.}\ \bibnamefont
  {Rabiner}},\ }\bibfield  {title} {\bibinfo {title} {A tutorial on hidden
  {Markov} models and selected applications in speech recognition},\ }\href
  {https://doi.org/10.1109/5.18626} {\bibfield  {journal} {\bibinfo  {journal}
  {Proc. IEEE}\ }\textbf {\bibinfo {volume} {77}},\ \bibinfo {pages} {257}
  (\bibinfo {year} {1989})}\BibitemShut {NoStop}%
\bibitem [{\citenamefont {Yu}(2010)}]{YU2010215}%
  \BibitemOpen
  \bibfield  {author} {\bibinfo {author} {\bibfnamefont {S.-Z.}\ \bibnamefont
  {Yu}},\ }\bibfield  {title} {\bibinfo {title} {Hidden semi-{Markov} models},\
  }\href {https://doi.org/10.1016/j.artint.2009.11.011} {\bibfield  {journal}
  {\bibinfo  {journal} {Artificial Intelligence}\ }\textbf {\bibinfo {volume}
  {174}},\ \bibinfo {pages} {215} (\bibinfo {year} {2010})}\BibitemShut
  {NoStop}%
\bibitem [{\citenamefont {Johnson}\ and\ \citenamefont
  {Willsky}(2013)}]{JohnsonWillsky2013}%
  \BibitemOpen
  \bibfield  {author} {\bibinfo {author} {\bibfnamefont {M.~J.}\ \bibnamefont
  {Johnson}}\ and\ \bibinfo {author} {\bibfnamefont {A.~S.}\ \bibnamefont
  {Willsky}},\ }\bibfield  {title} {\bibinfo {title} {Bayesian nonparametric
  hidden semi-{Markov} models},\ }\href
  {https://jmlr.org/papers/volume14/johnson13a/johnson13a.pdf} {\bibfield
  {journal} {\bibinfo  {journal} {Journal of Machine Learning Research}\
  }\textbf {\bibinfo {volume} {14}},\ \bibinfo {pages} {673} (\bibinfo {year}
  {2013})}\BibitemShut {NoStop}%
\bibitem [{\citenamefont {Lloyd}(1982)}]{Lloyd1982}%
  \BibitemOpen
  \bibfield  {author} {\bibinfo {author} {\bibfnamefont {S.~P.}\ \bibnamefont
  {Lloyd}},\ }\bibfield  {title} {\bibinfo {title} {Least squares quantization
  in {PCM}},\ }\href {https://doi.org/10.1109/TIT.1982.1056489} {\bibfield
  {journal} {\bibinfo  {journal} {IEEE Transactions on Information Theory}\
  }\textbf {\bibinfo {volume} {28}},\ \bibinfo {pages} {129} (\bibinfo {year}
  {1982})}\BibitemShut {NoStop}%
\bibitem [{\citenamefont {Cohen}(1960)}]{Cohen1960}%
  \BibitemOpen
  \bibfield  {author} {\bibinfo {author} {\bibfnamefont {J.}~\bibnamefont
  {Cohen}},\ }\bibfield  {title} {\bibinfo {title} {A coefficient of agreement
  for nominal scales},\ }\href {https://doi.org/10.1177/001316446002000104}
  {\bibfield  {journal} {\bibinfo  {journal} {Educational and Psychological
  Measurement}\ }\textbf {\bibinfo {volume} {20}},\ \bibinfo {pages} {37}
  (\bibinfo {year} {1960})}\BibitemShut {NoStop}%
\bibitem [{\citenamefont {Fisher}(1936)}]{Fisher1936}%
  \BibitemOpen
  \bibfield  {author} {\bibinfo {author} {\bibfnamefont {R.~A.}\ \bibnamefont
  {Fisher}},\ }\bibfield  {title} {\bibinfo {title} {The use of multiple
  measurements in taxonomic problems},\ }\href
  {https://doi.org/10.1111/j.1469-1809.1936.tb02137.x} {\bibfield  {journal}
  {\bibinfo  {journal} {Annals of Eugenics}\ }\textbf {\bibinfo {volume} {7}},\
  \bibinfo {pages} {179} (\bibinfo {year} {1936})}\BibitemShut {NoStop}%
\bibitem [{\citenamefont {Jolliffe}(2014)}]{Jolliffe}%
  \BibitemOpen
  \bibfield  {author} {\bibinfo {author} {\bibfnamefont {I.}~\bibnamefont
  {Jolliffe}},\ }\bibfield  {title} {\bibinfo {title} {Principal component
  analysis},\ }in\ \href {https://doi.org/10.1002/9781118445112.stat06472}
  {\emph {\bibinfo {booktitle} {Wiley StatsRef: Statistics Reference Online}}}\
  (\bibinfo  {publisher} {John Wiley \& Sons, Ltd},\ \bibinfo {year}
  {2014})\BibitemShut {NoStop}%
\bibitem [{\citenamefont {Dunn}(1961)}]{Dunn1961}%
  \BibitemOpen
  \bibfield  {author} {\bibinfo {author} {\bibfnamefont {O.~J.}\ \bibnamefont
  {Dunn}},\ }\bibfield  {title} {\bibinfo {title} {Multiple comparisons among
  means},\ }\href {https://doi.org/10.1080/01621459.1961.10482090} {\bibfield
  {journal} {\bibinfo  {journal} {Journal of the American Statistical
  Association}\ }\textbf {\bibinfo {volume} {56}},\ \bibinfo {pages} {52}
  (\bibinfo {year} {1961})}\BibitemShut {NoStop}%
\bibitem [{\citenamefont {Hoang}\ and\ \citenamefont
  {Baur}(2022)}]{HOANG2022106622}%
  \BibitemOpen
  \bibfield  {author} {\bibinfo {author} {\bibfnamefont {L.~T.}\ \bibnamefont
  {Hoang}}\ and\ \bibinfo {author} {\bibfnamefont {D.~G.}\ \bibnamefont
  {Baur}},\ }\bibfield  {title} {\bibinfo {title} {Loaded for bear: {Bitcoin}
  private wallets, exchange reserves and prices},\ }\href
  {https://doi.org/10.1016/j.jbankfin.2022.106622} {\bibfield  {journal}
  {\bibinfo  {journal} {Journal of Banking \& Finance}\ }\textbf {\bibinfo
  {volume} {144}},\ \bibinfo {pages} {106622} (\bibinfo {year}
  {2022})}\BibitemShut {NoStop}%
\bibitem [{\citenamefont {Bertucci}(2024)}]{BERTUCCI2024102338}%
  \BibitemOpen
  \bibfield  {author} {\bibinfo {author} {\bibfnamefont {L.}~\bibnamefont
  {Bertucci}},\ }\bibfield  {title} {\bibinfo {title} {{Bitcoin} {Ordinals}:
  Determinants and impact on total transaction fees},\ }\href
  {https://doi.org/10.1016/j.ribaf.2024.102338} {\bibfield  {journal} {\bibinfo
   {journal} {Research in International Business and Finance}\ }\textbf
  {\bibinfo {volume} {70}},\ \bibinfo {pages} {102338} (\bibinfo {year}
  {2024})}\BibitemShut {NoStop}%
\bibitem [{\citenamefont {Wiedenmann}\ and\ \citenamefont
  {Guettler}(2026)}]{wiedenmann2025ordinals}%
  \BibitemOpen
  \bibfield  {author} {\bibinfo {author} {\bibfnamefont {A.}~\bibnamefont
  {Wiedenmann}}\ and\ \bibinfo {author} {\bibfnamefont {A.}~\bibnamefont
  {Guettler}},\ }\bibfield  {title} {\bibinfo {title} {{Bitcoin} {Ordinals} and
  {Inscriptions}: An analysis of {Bitcoin}'s evolving network dynamics},\
  }\href {https://doi.org/10.31585/jbba-9-1-(1)2026} {\bibfield  {journal}
  {\bibinfo  {journal} {The Journal of the British Blockchain Association}\
  }\textbf {\bibinfo {volume} {9}},\ \bibinfo {pages} {1} (\bibinfo {year}
  {2026})}\BibitemShut {NoStop}%
\bibitem [{\citenamefont {Divakaruni}\ and\ \citenamefont
  {Zimmerman}(2023)}]{DIVAKARUNI2023103480}%
  \BibitemOpen
  \bibfield  {author} {\bibinfo {author} {\bibfnamefont {A.}~\bibnamefont
  {Divakaruni}}\ and\ \bibinfo {author} {\bibfnamefont {P.}~\bibnamefont
  {Zimmerman}},\ }\bibfield  {title} {\bibinfo {title} {The {Lightning
  Network}: Turning {Bitcoin} into money},\ }\href
  {https://doi.org/10.1016/j.frl.2022.103480} {\bibfield  {journal} {\bibinfo
  {journal} {Finance Research Letters}\ }\textbf {\bibinfo {volume} {52}},\
  \bibinfo {pages} {103480} (\bibinfo {year} {2023})}\BibitemShut {NoStop}%
\bibitem [{\citenamefont {Alexander}\ and\ \citenamefont
  {Heck}(2020)}]{ALEXANDER2020100776}%
  \BibitemOpen
  \bibfield  {author} {\bibinfo {author} {\bibfnamefont {C.}~\bibnamefont
  {Alexander}}\ and\ \bibinfo {author} {\bibfnamefont {D.~F.}\ \bibnamefont
  {Heck}},\ }\bibfield  {title} {\bibinfo {title} {Price discovery in
  {Bitcoin}: The impact of unregulated markets},\ }\href
  {https://doi.org/10.1016/j.jfs.2020.100776} {\bibfield  {journal} {\bibinfo
  {journal} {Journal of Financial Stability}\ }\textbf {\bibinfo {volume}
  {50}},\ \bibinfo {pages} {100776} (\bibinfo {year} {2020})}\BibitemShut
  {NoStop}%
\bibitem [{\citenamefont {Mohamad}(2025)}]{Mohamad2025}%
  \BibitemOpen
  \bibfield  {author} {\bibinfo {author} {\bibfnamefont {A.}~\bibnamefont
  {Mohamad}},\ }\bibfield  {title} {\bibinfo {title} {Do {Bitcoin} {ETFs} lead
  price discovery following their introduction in the {Bitcoin} market?},\
  }\href {https://doi.org/10.1007/s10614-025-10998-x} {\bibfield  {journal}
  {\bibinfo  {journal} {Computational Economics}\ }\textbf {\bibinfo {volume}
  {66}},\ \bibinfo {pages} {947} (\bibinfo {year} {2025})}\BibitemShut
  {NoStop}%
\bibitem [{\citenamefont {Granger}(1969)}]{Granger1969}%
  \BibitemOpen
  \bibfield  {author} {\bibinfo {author} {\bibfnamefont {C.~W.~J.}\
  \bibnamefont {Granger}},\ }\bibfield  {title} {\bibinfo {title}
  {Investigating causal relations by econometric models and cross-spectral
  methods},\ }\href {https://doi.org/10.2307/1912791} {\bibfield  {journal}
  {\bibinfo  {journal} {Econometrica}\ }\textbf {\bibinfo {volume} {37}},\
  \bibinfo {pages} {424} (\bibinfo {year} {1969})}\BibitemShut {NoStop}%
\bibitem [{\citenamefont {Pearl}(2009)}]{Pearl2009}%
  \BibitemOpen
  \bibfield  {author} {\bibinfo {author} {\bibfnamefont {J.}~\bibnamefont
  {Pearl}},\ }\href {https://doi.org/10.1017/CBO9780511803161} {\emph {\bibinfo
  {title} {Causality: Models, Reasoning, and Inference}}},\ \bibinfo {edition}
  {2nd}\ ed.\ (\bibinfo  {publisher} {Cambridge University Press},\ \bibinfo
  {address} {Cambridge},\ \bibinfo {year} {2009})\BibitemShut {NoStop}%
\bibitem [{\citenamefont {Sugihara}\ \emph {et~al.}(2012)\citenamefont
  {Sugihara}, \citenamefont {May}, \citenamefont {Ye}, \citenamefont {Hsieh},
  \citenamefont {Deyle}, \citenamefont {Fogarty},\ and\ \citenamefont
  {Munch}}]{Sugihara2012}%
  \BibitemOpen
  \bibfield  {author} {\bibinfo {author} {\bibfnamefont {G.}~\bibnamefont
  {Sugihara}}, \bibinfo {author} {\bibfnamefont {R.}~\bibnamefont {May}},
  \bibinfo {author} {\bibfnamefont {H.}~\bibnamefont {Ye}}, \bibinfo {author}
  {\bibfnamefont {C.-h.}\ \bibnamefont {Hsieh}}, \bibinfo {author}
  {\bibfnamefont {E.}~\bibnamefont {Deyle}}, \bibinfo {author} {\bibfnamefont
  {M.}~\bibnamefont {Fogarty}},\ and\ \bibinfo {author} {\bibfnamefont
  {S.}~\bibnamefont {Munch}},\ }\bibfield  {title} {\bibinfo {title} {Detecting
  causality in complex ecosystems},\ }\href
  {https://doi.org/10.1126/science.1227079} {\bibfield  {journal} {\bibinfo
  {journal} {Science}\ }\textbf {\bibinfo {volume} {338}},\ \bibinfo {pages}
  {496} (\bibinfo {year} {2012})}\BibitemShut {NoStop}%
\bibitem [{\citenamefont {Di~Antonio}\ and\ \citenamefont
  {Vinci}(2025)}]{DiAntonio2025Causality}%
  \BibitemOpen
  \bibfield  {author} {\bibinfo {author} {\bibfnamefont {G.}~\bibnamefont
  {Di~Antonio}}\ and\ \bibinfo {author} {\bibfnamefont {G.~V.}\ \bibnamefont
  {Vinci}},\ }\bibfield  {title} {\bibinfo {title} {Nonlinear correlations
  underlie linear response and causality},\ }\href
  {https://doi.org/10.1103/vjvt-gdn1} {\bibfield  {journal} {\bibinfo
  {journal} {Physical Review Research}\ }\textbf {\bibinfo {volume} {7}},\
  \bibinfo {pages} {L042029} (\bibinfo {year} {2025})}\BibitemShut {NoStop}%
\end{thebibliography}%

\clearpage
\pagestyle{empty}
\onecolumngrid
\section*{Supplementary materials}

\setcounter{figure}{0}
\setcounter{table}{0}
\renewcommand{\thefigure}{S\arabic{figure}}
\renewcommand{\thetable}{S\arabic{table}}

\begin{figure*}[ht!]
\centering
\includegraphics[width=\textwidth]{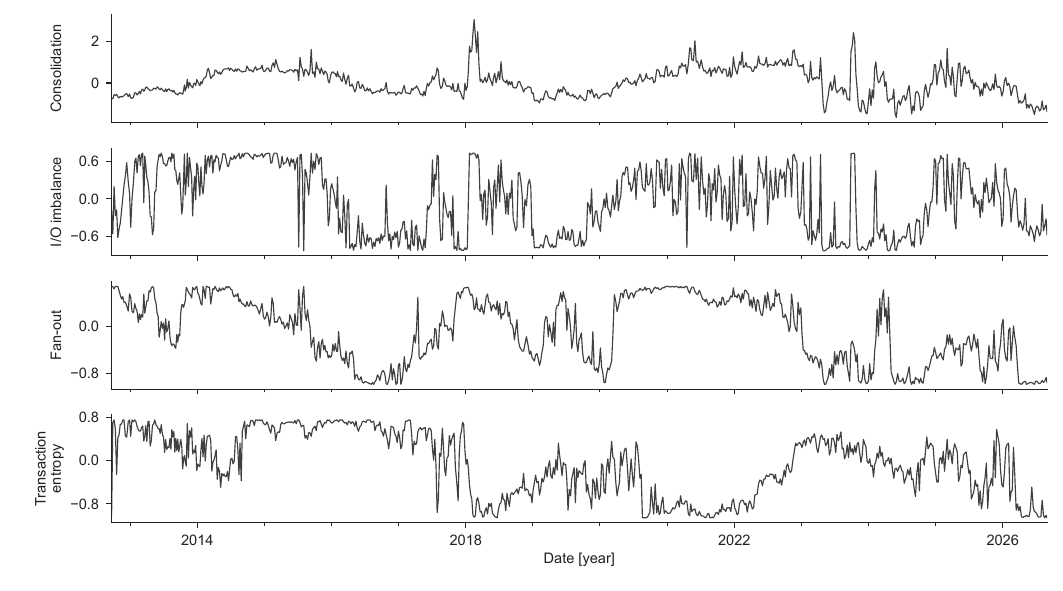}
\caption{\textbf{Final on-chain features used by the HSMM.}
Time series of the four model inputs over the 760 weeks used by the model.
Input--output imbalance, fan-out and transaction entropy are trailing percentile ranks. All four are then centred on their full-sample medians and divided by their interquartile ranges.}
\label{fig:features}
\end{figure*}

\begin{figure*}[ht!]
\centering
\includegraphics[width=\textwidth]{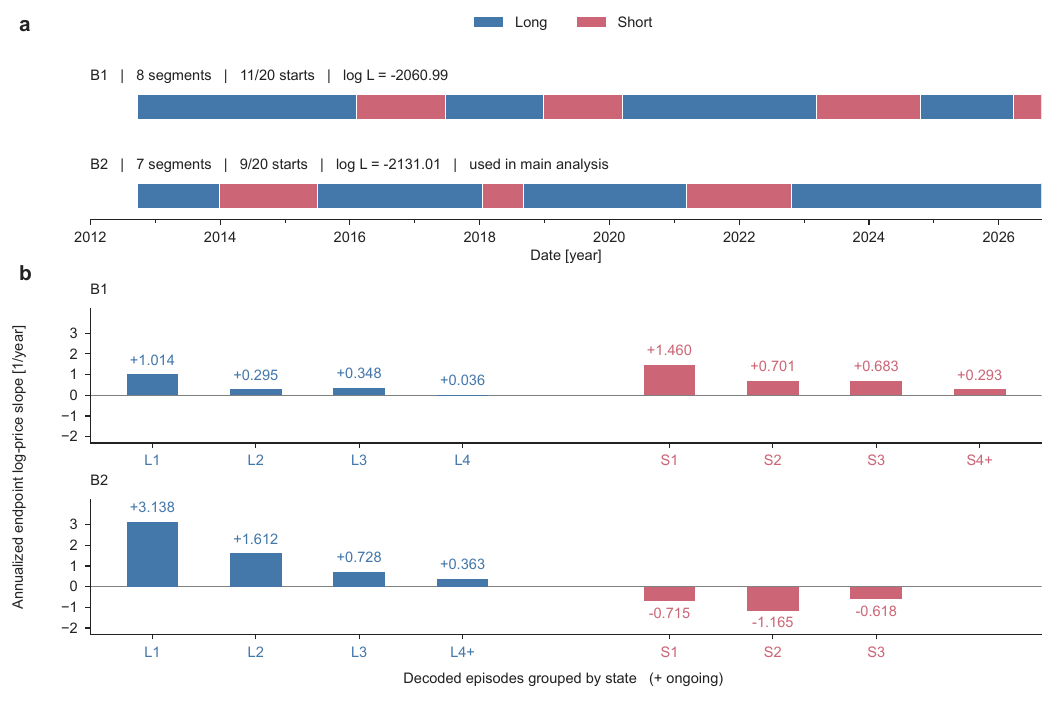}
\caption{\textbf{Dependence of the HSMM solution on initialisation.}
\textbf{a,} Twenty model initialisations converge to two distinct decoded
state sequences. Basin B1 is recovered in 11 of 20 starts and contains
eight episodes, while basin B2 is recovered in 9 of 20 starts and contains
seven episodes.
\textbf{b,} Annualised endpoint log-price slopes for the decoded episodes
of each solution, calculated as in Fig.~\ref{fig:states}c and grouped by
latent state. In B2, all Long episodes have positive slopes and all Short
episodes have negative slopes, whereas both states exhibit positive slopes
in B1. Although B1 has the higher likelihood, B2 shows a clearer
correspondence between the inferred blockchain regimes and historical
market direction.}
\label{fig:solutions}
\end{figure*}

\begin{figure*}[ht!]
\centering
\includegraphics[width=\textwidth]{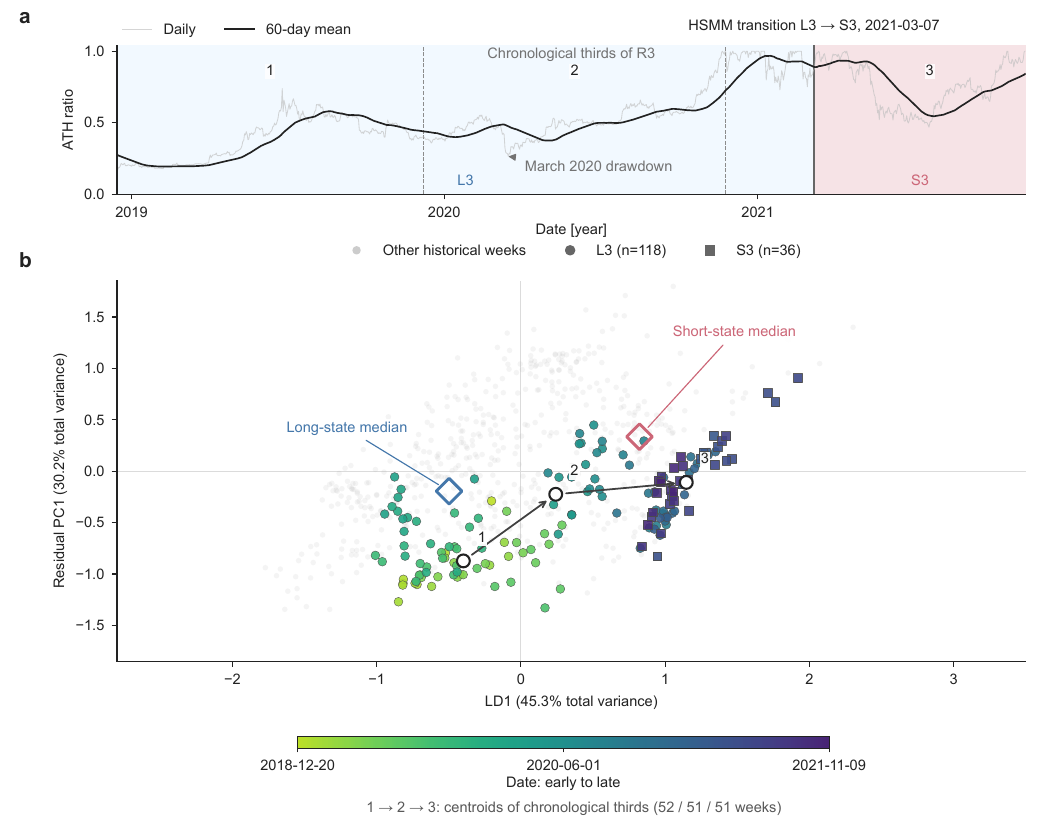}
\caption{\textbf{Within-phase evolution of on-chain transaction structure during \(\mathbf{R3}\).}
\textbf{a,} Bitcoin all-time-high ratio during the externally defined
\(R3\) Rising phase (2018-12-15 to 2021-11-10). The daily ratio is shown
in grey and its trailing 60-day mean in black. Although \(R3\) forms one
broad Rising phase, it contains substantial internal variation, including
the pronounced March 2020 drawdown. Dashed vertical lines divide the
interval into three consecutive equal-count chronological partitions,
labelled \(1\)--\(3\). The solid vertical line marks the HSMM transition
from \(L3\) to \(S3\) on 2021-03-07.
\textbf{b,} The same \(R3\) observations projected onto the
LD1--Residual-PC1 coordinates of Fig.~\ref{fig:nonstationarity}d.
The 118 weeks decoded as \(L3\) are shown as circles and the 36 weeks
decoded as \(S3\) as squares. Colour indicates time from early to
late. Light-grey points show the remaining historical observations, while
diamonds mark the pooled Long- and Short-state medians. Open circles
labelled \(1\)--\(3\) show the centroids of the three chronological
partitions defined in panel \textbf{a}, with arrows indicating their
temporal ordering. The distribution shifts through \(R3\) towards the
Short-associated side of LD1, while the blockchain transition to \(S3\)
occurs before the end of the price-defined Rising phase.}
\label{fig:R3}
\end{figure*}

\end{document}